\documentclass[aps,pre,twocolumn,footinbib,superscriptaddress,floatfix]{revtex4-2}

\usepackage{graphicx}
\usepackage{dblfloatfix}
\usepackage{amssymb}
\usepackage{amsmath}
\usepackage{times}
\usepackage{bm}
\usepackage{xcolor}

\begin{document}
	
\title{Roughening of the Anharmonic Elastic Interface in Correlated Random Media}

\author{Alejandro Al\'es}
\affiliation{Instituto de F\'isica de Materiales Tandil (IFIMAT), 
	Facultad de Ciencias Exactas, Universidad Nacional 
	del Centro de la Provincia de Buenos Aires (UNCPBA), Pinto 399, 7000 Tandil, Argentina}

\affiliation{Consejo Nacional de Investigaciones Cient\'ificas y T\'ecnicas (CONICET), Argentina}

\author{Juan M. L\'opez}
\email{lopez@ifca.unican.es}
\affiliation{Instituto de F\'isica de Cantabria (IFCA), CSIC-Universidad de
	Cantabria, 39005 Santander, Spain}

\date{\today}

\begin{abstract}
We study the roughening properties of the anharmonic elastic interface in the presence of temporally correlated noise. The model can be seen as a generalization of the anharmonic Larkin model, recently introduced by Purrello, Iguain, and Kolton [Phys. Rev. E {\bf 99}, 032105 (2019)], to investigate the effect of higher-order corrections to linear elasticity in the fate of interfaces. We find analytical expressions for the critical exponents as a function of the anharmonicity index $n$, the noise correlator range $\theta \in[0,1/2]$, and dimension $d$. In $d=1$ we find that the interface becomes faceted and exhibits anomalous scaling for $\theta > 1/4$ for any degree of anharmonicity $n > 1$. Analytical expressions for the anomalous exponents $\alpha_\mathrm{loc}$ and $\kappa$ are obtained and compared with a numerical integration of the model. Our theoretical results show that anomalous roughening cannot exist for this model in dimensions $d > 1$. 
\end{abstract}


\maketitle

\section{Introduction}
The dynamics of elastic interfaces and strings in disordered media is a ubiquitous topic in statistical mechanics and condensed matter physics~\cite{Barabasi.Stanley1995_book,Krug1997,Fisher1998,Kardar_1998}. From a fundamental point of view this problem offers interesting examples of critical dynamics, universality, and scale-free behavior in the presence of disorder/noise, often far from equilibrium~\cite{Nattermann1992,Leschhorn1997,Jensen_1995,Rosso.Krauth_2001,Duemmer.Krauth_2005,Rosso.etal_2003,Rosso.Krauth_2002,Chauve.etal_2001,Kolton.etal_2005a}. Applications include the roughening of magnetic domain walls in disordered materials~\cite{Bruinsma.Aeppli_1984,Halpin-Healy_1989,Huse.Henley_1985}, magnetic flux lines in type-II superconductors~\cite{Blatter1994,Ertas_1996,Nattermann2000}, fracture cracks propagation~\cite{Ramanathan.etal_1997,Bouchaud1993,Bouchaud_1997,Ponson2017}, fluid imbibition in porous media~\cite{Dube1999,Hernandez-MachadoA.2001,Alava.etal_2004,Rost.etal_2007,Pradas2007,Pradas2009}, among many others. Disorder can locally pin the interface, while the external driving force $f$ pushes the interface forward. Collective interface motion can occur whenever $f$ is above a threshold value $f_\mathrm{c}$. At zero temperature this leads to a critical depinning transition at $f=f_\mathrm{c}$, when elastic forces provide the enough nearest-neighbor interaction for collective motion to appear. For driving forces below the critical threshold the interface stops moving. In contrast, at finite temperatures there are interesting effects like glassiness and creeping due to the fact that the interface can move, albeit extremely slowly, even below the critical forcing due to thermal fluctuations~\cite{Nattermann.etal_1990,Chauve2000,Kolton2009,de_la_Lama_2009,Purrello2017}.

At zero temperature the relaxational dynamics of an interface $h(\mathbf{x},t)$ at time $t$ and substrate position $\mathbf{x} \in \Re^d$ ,  with the harmonic elastic Hamiltonian functional ${\cal H}[h] = (\nu/2) \int d^d\mathbf{x} (\mathbf{\nabla}h)^2$, and subject to local pinning forces $\eta(\mathbf{x},y)$ is given by
\begin{equation}
\frac{\partial h}{\partial t} = - \frac{\delta \cal H}{\delta h(\mathbf{x},t)} + f + \eta(\mathbf{x},h),
\label{eq:elastic}
\end{equation}
where $f$ is the driving force and $\eta(\mathbf{x},y)$ is a static random field that represents the disordered medium in which the interface moves through. The details of the disorder are irrelevant for the interface behavior and it suffices to specify the first two moments. The disorder has zero mean, $\langle \overline{\eta(\mathbf{x},y)} \rangle = 0$, where average both in $\mathbf{x}$  and over disorder realizations, $\langle \cdots \rangle$, are implied and the disorder correlation is typically short-ranged $\langle\overline{\eta(\mathbf{x},y) \eta(\mathbf{x}',y')}\rangle = \sigma^2 \delta^d(\mathbf{x} - \mathbf{x}') \Delta_\xi(y - y')$, where $\Delta_\xi(y)$ is an even function with a short spatial range $\xi$~\cite{Nattermann1992,Leschhorn1997}. 

The interface dynamics resulting from (\ref{eq:elastic}) is the so-called quenched Edwards-Wilkinson (QEW) equation:
\begin{equation}
\frac{\partial h}{\partial t} = \nu \mathbf{\nabla}^2 h + f + \eta(\mathbf{x},h),
\label{eq:qew}
\end{equation}
and represents the universality class of models that describe the behavior of an elastic interface in the harmonic approximation moving through a disordered medium.

Starting from a flat initial state Eq.~(\ref{eq:qew}) leads to scale-invariant behavior, as can be realized by calculating the interface correlations $C(\ell,t) = \langle\overline{[h(\mathbf{x},t) - h(\mathbf{x} + \mathbf{l},t)]^2}\rangle^{1/2} = \ell^{\alpha} {\cal G}(\ell/t^{1/z})$, where $\ell = |\mathbf{l}|$, or the interface global width in a system of size $L$,
\begin{equation}
W (L,t) = \langle \overline{[h(\mathbf{x},t) - \langle h \rangle)]^2} \rangle^{1/2} = L^\alpha {\cal F}(L/t^{1/z}),
\label{eq:global_W}
\end{equation}
where $\alpha$ and $z$ are the roughness and dynamic exponents, respectively, which characterize the critical properties of the interface, and ${\cal F}$ is a scaling function that becomes constant, ${\cal F} (u) \sim \mathrm{const.}$, for $u \ll 1$ and scales as ${\cal F}(u) \sim u^{-\alpha}$ for $u \gg 1$. A similar scaling behavior is expected for ${\cal G}(\ell/t^{1/z})$. The actual value of the critical exponents $\alpha$ and $z$ changes across the depinning phase transition point at $f=f_c$~\cite{Makse_1995,Amaral1995}.

Analytical treatment of the QEW model is difficult due to the nonlinear dependence implicit in the pinning force and requires methods, like the functional renormalization group~\cite{Nattermann1992,Leschhorn1997}, that treat the problem perturbatively in the disorder. At lowest-order, this consists in neglecting the dependence on $y$ in the pinning force and, hence, replace the disordered field by a columnar disorder. By doing so one arrives at the so-called Larkin model~\cite{Larkin1979,Blatter1994,Nattermann2000}:
\begin{equation}
\frac{\partial h}{\partial t} = \nu \mathbf{\nabla}^2 h + f + \eta(\mathbf{x}),
\label{eq:larkin}
\end{equation}
where the driving force can be eliminated by choosing a co-moving reference frame, $h \to h + ft$, and will be taken as zero in this paper. The Larkin model is a crude, lowest-order, approximation to the QEW problem that neglects important correlations in the growth direction implied by the dependence of the pinning force $\eta(\mathbf{x},h)$ on the interface state $h(\mathbf{x},t)$. Nonetheless, this approximation is expected to be interesting to describe the pinned phase, where the interface halts.  Columnar disorder is also relevant in other contexts like the transport and localization properties of particles in trapping and amplifying disordered media~\cite{Zhang1986,Engel1987,Tao1988,Guyer1990,Szendro2007a}. Therefore, Eq.~(\ref{eq:larkin}) has interest in its own. 

Within the Larkin approximation the problem becomes linear and the critical exponents can be found exactly, either by solving  (\ref{eq:larkin}) explicitly by Green propagator methods or by simple dimensional analysis, to obtain $\alpha = (4-d)/2$ and $z = 2$. We immediately note that the one-dimensional string becomes super-rough with $\alpha = 3/2>1$, which implies some anomalous scaling properties of the local interface fluctuations (see Ref.~\cite{Ramasco.etal_2000} for the general theory of anomalous surface scaling). In particular, super-rough behavior implies a local interface scaling distinctively different from that of the global interface fluctuations in (\ref{eq:global_W}). Indeed, one finds $C(\ell,t) \sim t^{\kappa} \, \ell$ with $\kappa = (\alpha -1)/z$, for intermediate times  $\ell \ll t^{1/z} \ll L$, while in the stationary regime, $t^{1/z} \gg L$, we have $C(\ell, t\to\infty) \sim \ell \, L^{\alpha-1}$~\cite{Ramasco.etal_2000}. Interestingly,  super-rough interfaces also appear in the QEW model itself, Eq.(\ref{eq:qew}), where one finds $\alpha = 1.25$ in $d=1$ at the depinning phase transition point $f=f_c$~\cite{Leschhorn1993,Jensen_1995,Chauve.etal_2001,Rosso.Krauth_2001,Rosso.etal_2003}. 

An important consequence of having $\alpha > 1$ is that, within the elastic description, local elongations in the growth direction become unbounded in the thermodynamic limit, $W(L)/L \sim L^{\alpha -1} \to \infty$ as $L \to \infty$.  Rosso and Krauth~\cite{Rosso2001} proposed this is physically unacceptable and just indicates that higher-order anharmonic corrections should be taken into account to cure this issue. By means of numerical simulations they showed~\cite{Rosso2001} that, indeed, anharmonic corrections cure the QEW model by leading to a roughness exponent $\alpha = 0.63 < 1$ in $d=1$, remarkably, a value very close to that of the quenched Kardar-Parisi-Zhang universality (QKPZ) class at depinning~\cite{Tang.Leschhorn_1992,Buldyrev1992,Amaral1995}. 

Inspired by this idea, Purrello {\it et al.}~\cite{Purrello2019} very recently studied the $d=1$ Larkin model with anharmonic corrections by adding to the evolution equation (\ref{eq:larkin}) a generic term of the form $c_{2n} \nabla(\nabla h)^{2n-1}$, herewith called anharmonic Larkin model (ALM):  
\begin{eqnarray}
\frac{\partial h}{\partial t} = \nu \mathbf{\nabla}^2 h + c_{2n}\mathbf{\nabla} (\mathbf{\nabla} h)^{2n-1} + \eta(\mathbf{x}),
\label{eq:ALM}
\end{eqnarray}
with the coefficients $c_{2n} >0$ and $n = 2, 3, 4, \dots$. These anharmonic terms are Hamiltonian and can be interpreted as higher-order corrections to elasticity (see Sec.~\ref{sec:model}). For large $n$ the net effect of such a term would be to constrain local elongations and, therefore, Purrello {\it et al.}~\cite{Purrello2019} asked the question whether the ALM could lead to the same exponents as those of KPZ with temporally correlated noise for some value of the nonlinear index $n$. However, in their study of the ALM, they found an anomalous {\em faceted} scaling, with the spectral roughness exponent $\alpha_\mathrm{s}$ satisfying $\alpha_\mathrm{s} > \alpha > 1$ for any finite $n > 1$, invalidating the usual single-exponent scaling for two-point correlation functions, and the small gradient approximation of the elastic energy density in the thermodynamic limit (see Sec. \ref{sec:tools} for details on faceted roughening and the meaning of the spectral roughness exponent $\alpha_\mathrm{s}$).  

In this paper we study the roughening properties of the anharmonic elastic interface in the presence of long-time correlated noise. The model can be seen as a generalization of the ALM where one replaces the columnar disorder, $\eta(\mathbf{x})$ in Eq.~(\ref{eq:ALM}) by a temporally correlated noise $\eta(\mathbf{x},t)$, with $\langle \eta(\mathbf{x},t) \eta(\mathbf{x}',t')\rangle = 2D\, \delta^d(\mathbf{x} - \mathbf{x}') \, |t-t'|^{2\theta-1}$. The advantage of this modification is that it allows to accommodate more general settings, where the columnar disorder is a too strong approximation and, at the same time, it provides a solid ground to investigate the effect of anharmonic correction terms to elasticity in the presence of time correlated noise. Since some of the model couplings do not renormalize, several exponents can be well approximated by a scaling theory. We find analytical expressions for the critical exponents $\alpha$ and $z$ as a function of the anharmonicity index $n$, noise exponent $\theta$, and dimension $d$. For $d = 1$ we find that the interface is faceted ($\alpha_s \neq \alpha$) and exhibits anomalous scaling for $\theta > 1/4$ for any $n>1$. This includes, in particular, the columnar case ($\theta = 1/2$) recently studied by Purrello {\it et al.}~\cite{Purrello2019}. By studying the scaling behavior of the local slope field $\Upsilon = \nabla h$ we also obtain an analytical expression for the anomalous time exponent $\kappa$ and show that anomalous scaling corresponds to the region of $\theta$ values such that $\kappa > 0$, in agreement with existing theoretical results about the origin of anomalous kinetic roughening~\cite{Lopez_1999}. We conclude that anomalous roughening cannot exist for dimensions $d > 1$ in the ALM (even with correlated noise) and the interface exhibits standard single-exponent scaling, $\alpha = \alpha_\mathrm{loc} = \alpha_\mathrm{s}$ in that case. Our theoretical results are in good agreement with the numerical simulations of the model in $d=1$ that we carried out and also present here.

The paper is organized as follows. In Sec.~\ref{sec:tools} we introduce the mathematical quantities, scaling functions and exponents that are used to characterize interface kinetic roughening, including some basics on anomalous scaling behavior and how to detect it. Sec.~\ref{sec:model} is devoted to discuss the ALM with temporal correlations and we introduce a scaling theory for the roughening exponents. In Sec.~\ref{sec:numeric} we present our numerical results and comparison with theory. Finally, we conclude with a discussion of the results in Sec.~\ref{sec:discussion}

\section{Anomalous kinetic roughening: Scaling functions and exponents}
\label{sec:tools}
In this section we introduce some of the tools that we will be using hereafter to study anomalously roughened interfaces, particularly for the case of rough facets. The statistical properties of rough scale-invariant interfaces remain  unchanged after re-scaling of space and time according to the transformation $h(\mathbf{x},t) \to b^{\alpha} h(b \, \mathbf{x}, b^{1/z} \, t)$, for any scaling factor $b>1$ and some specific value of the critical  exponents $\alpha$ and $z$~\cite{Barabasi.Stanley1995_book,Krug1997}. In practice, the exponents may be obtained from correlation functions like the local interface correlations $C(\ell,t)$ and the global interface width $W(L,t)$ already discussed above. However, the most general description of the interface scaling properties is best achieved by using the structure factor $S(k,t) = \langle \widehat{h}(\mathbf{k},t) \widehat{h}(-\mathbf{k},t) \rangle$, where $\widehat{h}(\mathbf{k},t) \equiv \int d^d\mathbf{x} \; h(\mathbf{x},t) \exp(-i \mathbf{k} \cdot \mathbf{x})$ 
is the Fourier transform of the interface profile $h(\mathbf{x},t)$, and $k = \vert \mathbf{k} \vert$. For kinetically roughened interfaces in dimension $d$ we expect 
\begin{equation}
S(k,t) = k^{-(2\alpha + d)} s(kt^{1/z}),
\label{eq:S}
\end{equation}
where the most general scaling function, consistent with scale-invariant dynamics, is given by~\cite{Ramasco.etal_2000}
\begin{equation}
s(u) \sim  
\left\{ \begin{array}{lcl}
u^{2(\alpha-\alpha_s}) & {\rm if} &  u \gg 1\\
u^{2\alpha + d} & {\rm if} &  u \ll 1
\end{array}
\right. ,
\label{eq:generic}
\end{equation}
with $\alpha$ being the {\em global} roughness exponent and $\alpha_s$ the so-called {\em spectral} roughness exponent~\cite{Ramasco.etal_2000}. Standard single-exponent scaling corresponds to $\alpha_s = \alpha < 1$. Remarkably, other situations can be described within this generic scaling framework, including super-roughening and intrinsic anomalous scaling, that may appear depending on the values of $\alpha_s$ and $\alpha$~\cite{Ramasco.etal_2000}. For faceted interfaces, the case of interest for us here, one has $\alpha_s > 1$ and $\alpha \neq \alpha_s$ so that two independent roughening exponents are actually needed to completely describe the scaling properties of the interface~\cite{Ramasco.etal_2000}. We stress here that only the structure factor $S(k,t)$ allows us to obtain the distinctively characteristic spectral roughness exponent, $\alpha_s$, typical of faceted growing interfaces, since this exponent leaves no direct trace in the local interface correlation $C(\ell,t)$ nor the global width $W(L,t)$~\cite{Ramasco.etal_2000}. Focusing on faceted anomalous scaling, one can calculate from Eqs.~(\ref{eq:S}) and (\ref{eq:generic}) the correlator { \bf $C(\ell,t) = \langle \overline{[h(\mathbf{x},t) - h(\mathbf{x}+\mathbf{l},t)]^2} \rangle \propto \int d^d\mathbf{k} [1 -\cos(\mathbf{k}\cdot\mathbf{l})] S(\mathbf{k},t)$ } to obtain $C(\ell,t) \sim \ell^{\alpha_\mathrm{loc}} \, t^\kappa$ for intermediate times $\ell < t^{1/z} < L$, where $\kappa = (\alpha - \alpha_\mathrm{loc})/z$. While the local interface fluctuations only saturate at times $t \gg L^z$, when they become time independent, and one finds $C(\ell,t) \sim \ell^{\alpha_\mathrm{loc}} \, L^{\alpha - \alpha_\mathrm{loc}}$. Also, and very importantly, for faceted interfaces one gets a fixed value for $\alpha_\mathrm{loc} = 1$ (see Ref.~\cite{Ramasco.etal_2000} for details). 

\section{Model and scaling theory}
\label{sec:model}
In this paper we shall study the roughening properties of the anharmonic elastic interface driven by noisy fluctuations with long temporal correlations. The time evolution of the interface $h(\mathbf{x},t)$ in dimension $d$ is described by  
\begin{eqnarray}
\frac{\partial h}{\partial t} = \nu \mathbf{\nabla}^2 h + c_{2n}\mathbf{\nabla} (\mathbf{\nabla} h)^{2n-1} + \eta(\mathbf{x},t),
\label{eq:our_model}
\end{eqnarray}
where the coupling coefficients $c_{2n}$ are positive, $n = 2, 3, \cdots$, and the noise has correlations given by 
\begin{equation}
\langle \eta(\mathbf{x},t) \eta(\mathbf{x}',t')\rangle = 2D \, \delta^d(\mathbf{x} - \mathbf{x}') \, |t-t'|^{2\theta-1},
\label{eq:noise}
\end{equation}
where the index $\theta \in [0,1/2]$ controls the extent of the temporal correlations. The ALM equation~(\ref{eq:ALM}) is recovered in the limit $\theta \to 1/2$, when the noise becomes a (spatially uncorrelated) static columnar disorder term $\eta(\mathbf{x})$.
 
This model can be seen as the $T=0$ Langevin relaxational dynamics, via Eq.(\ref{eq:elastic}), for a manifold that obeys the elastic Hamiltonian functional ${\cal H}[h] = (\nu/2) \int d^d\mathbf{x} \sqrt{1 + (\mathbf{\nabla}h)^2}$, after a Taylor expansion $\sqrt{1 + (\mathbf{\nabla}h)^2} = \sum_{n=1}^\infty a_n (\mathbf{\nabla} h)^{2n}$. At lowest-order, this expansion gives the usual Laplacian term, while higher-order terms of the form $\nabla (\nabla h)^{2n-1}$ should be relevant if $\mathbf{\nabla} h$ is not small, which is the case for the model considered here~\cite{footnote}. 

It is important to start this study with an analysis of the relevance of a nonlinear term like $c_{2n}\nabla (\nabla h)^{2n-1}$  as compared with the Laplacian term $\nu \mathbf{\nabla}^2 h$ in the long wavelengths limit. Consider we scale distances, $x \to bx$, by a factor $b>1$, while the scale-invariant interface should rescale as $h \to b^\alpha h$. Accordingly, the Laplacian term will become $b^{\alpha-2} \nabla^2 h$, and the nonlinear term $c_{2n} b^{2n(\alpha-1)-\alpha} \nabla (\nabla h)^{2n-1}$. At long wavelengths, $b \to \infty$, the nonlinear term becomes larger than the Laplacian one if and only if $\alpha -2 < 2n(\alpha-1)-\alpha$. This immediately leads to the condition $\alpha > 1$ for any $n \geq 2$ in any dimension. 
On the contrary, for $\alpha < 1$ the nonlinear term is irrelevant as far as the asymptotic scaling is concerned. This observations will be important later on in order to find the correct scaling exponents as a function of the noise correlation index $\theta$.

\subsection{Theoretical results}
We now construct a simple, but exact, scaling theory based on dimensional analysis that gives the critical exponent values as a function of the nonlinearity index $n \geq 2$ and the noise exponent $\theta$. The reason why dimensional analysis gives the exact scaling exponents is that none of the equation coupling constants renormalizes after scaling. By imposing the scale invariance of the interface field $h \to b^\alpha h$ as spatial coordinates, $x \to b x$ and time, $t \to b^z t$, are scaled by a factor $b > 1$, the time evolution equation (\ref{eq:our_model}) becomes
\begin{eqnarray}
b^{\alpha - z} \frac{\partial h}{\partial t} &=&  \nu b^{\alpha - 2} \mathbf{\nabla}^2 h + c_{2n} b^{(2n-1)\alpha - 2n} \mathbf{\nabla} (\mathbf{\nabla} h)^{2n-1} +  \nonumber \\ 
&+& b^{-d/2+ z(\theta-1/2)} \eta(x,t), 
\label{eq:our_model_rescaled}
\end{eqnarray}
where we have taken into account the noise correlator (\ref{eq:noise}) to rescale the stochastic term. We now assume the $n$-nonlinearity is dominant in the long wavelength limit and neglect the Laplacian term. By imposing invariance of the equation of motion (\ref{eq:our_model_rescaled}) for any $b > 1$ we get to the scaling relations
\[ 2\alpha (n-1) +z = 2n \]
\[ -d/2 + z (\theta+1/2) = \alpha,  \]
immediately implying the exact critical exponents in dimension $d$
\begin{eqnarray}
\alpha(n,\theta) &=& \frac{2n(2\theta+1)-d}{2n(2\theta+1)-4\theta}
\label{eq:alpha} \\
z(n,\theta) &=& \frac{n(d+2)-d}{n(2\theta+1)-2\theta}.
\label{eq:z}
\end{eqnarray}
These exponents will be valid as far as our assumption, {\it i.e.}, the $n$-nonlinearity is more relevant than the Laplacian term, is applicable. This is equivalent to the condition $\alpha(n,\theta) > 1$ (as discussed above) and leads, from (\ref{eq:alpha}), to a dimension-dependent bound condition on the noise index as $\theta > \theta_\mathrm{c} = d/4$ for the validity of (\ref{eq:alpha}) and (\ref{eq:z}). Note that $\theta_\mathrm{c} = 1/2$ for $d=2$ so, in reality, there is no anharmonic dominated regime for $d \ge 2$. 

Now, for $\theta < \theta_\mathrm{c} = d/4$ we have $\alpha(n,\theta) < 1$ and scaling relations (\ref{eq:alpha}) and (\ref{eq:z}) have to be modified, since the Laplacian term is then more relevant at long wavelengths than the nonlinearity for any $ n \geq 2$. This is actually the case for any $d >1$ and for $\theta < 1/4$ in $d=1$. In this case, neglecting the nonlinearity, we find
\begin{equation}
\alpha_\mathrm{EW}(\theta) = 2\theta + 1 -\frac{d}{2}\;\;\; \mathrm{and} \;\;\; z_\mathrm{EW}(\theta) = 2,
\label{eq:EW_exponents}
\end{equation} 
for $\theta < \theta_\mathrm{c} = d/4$. We use the label 'EW' to emphasize that these are the exponents in the Laplacian (harmonic elasticity) regime. These, indeed, correspond to the exact exponents for the purely Laplacian model, also known as the Edwards-Wilkinson equation, in the case of time correlated noise~\cite{Ales2020}.

We next calculate the corresponding local anomalous exponents $\alpha_\mathrm{loc}$ and $\kappa$ by direct computation using dimensional analysis as before. To do so we can resort to the theory developed in Ref.~\cite{Lopez_1999}, which allows us to compute the anomalous scaling exponents of the interface from the dynamics of the local slope field $\Upsilon(\mathbf{x},t) \equiv \mathbf{\nabla} h(x,t)$. The central point is to notice that anomalous scaling of any sort (intrinsic, super-roughening, or faceted) appears when the average local interface slope fluctuations grow in time, $\langle (\overline{\mathbf{\nabla} h})^2 \rangle^{1/2} \sim t^\kappa$, with some $\kappa > 0$. This new characteristic scale changes the scaling of the local interface fluctuations and leads to a local roughness exponent $\alpha_\mathrm{loc} = \alpha - \kappa z < \alpha$~\cite{Lopez_1999}. This allows, in principle, to determine the anomalous exponents $\kappa$ and $\alpha_\mathrm{loc}$ from the roughening properties of the local interface slope field $\Upsilon(\mathbf{x},t)$, which satisfies the equation of motion
\begin{eqnarray}
\frac{\partial \Upsilon}{\partial t} = \nu \mathbf{\nabla}^2 \Upsilon + c_{2n}\mathbf{\nabla}^2 (\Upsilon)^{2n-1} + \eta_c(\mathbf{x},t),
\label{eq:slopes}
\end{eqnarray}
where the noise is now conserved with correlation
\begin{equation*}
\langle \eta_c(\mathbf{x},t) \eta_c(\mathbf{x}',t')\rangle = -2D \; \mathbf{\nabla}^2\delta^d(\mathbf{x} - \mathbf{x}') \; |t-t'|^{2\theta-1}.
\end{equation*}
After rescaling of space $x \to b x$ and time $t \to b^{\tilde{z}} t$ we look for scale invariant solutions by rescaling the field by $\Upsilon \to b^{\tilde{\alpha}} \Upsilon$, where $\tilde{\alpha}$ and $\tilde{z}$ refer to the critical exponents of the interface slope field. Note that the anomalous time exponent corresponds to $\kappa = \tilde{\alpha}/\tilde{z}$. For simplicity of presentation we focus here on the $d=1$ case, where anomalies are expected. Applying the scale transformation to Eq.~(\ref{eq:slopes}) and equating terms we arrive at the exponents 
\begin{eqnarray}
\tilde{\alpha}(n,\theta) &=&  \frac{2 \theta -1/2}{\theta (2n-2) + n} \nonumber \\
\tilde{z}(n,\theta) &=&  \frac{3 n -1 }{\theta (2n-2) + n},
\label{eq:tilde_exp}
\end{eqnarray}
in $d=1$ when the Laplacian term is neglected. One can observe that above the threshold $\theta_\mathrm{c} = 1/4$ one has $\tilde{\alpha} > 0$ and the slope field becomes rough itself, marking the existence of anomalous behavior~\cite{Lopez_1999}. Using Eqs.~(\ref{eq:tilde_exp}) we find in $d=1$
\begin{equation}
\kappa(n,\theta) \equiv  \frac{\tilde{\alpha}}{\tilde{z}} = \frac{2\theta - 1/2}{3n-1},
\label{eq:kappa}
\end{equation}
for $\theta > \theta_\mathrm{c} = 1/4$ and $\kappa = 0$ otherwise.
This immediately gives the local roughness exponent, which is independent of the anharmonicity degree $n$,
\begin{equation}
\alpha_\mathrm{loc}(\theta)= \alpha - \kappa z = \left\{
\begin{array}{lcl}
2\theta + 1/2 \hspace{1.5mm} & \mathrm{if} & \theta \le 1/4  \\
1 & \mathrm{if} & \theta \ge 1/4
\end{array}
\right.
\label{eq:alpha_loc}     
\end{equation}
in $d=1$. For $d>1$, when the Laplacian always dominates over the nonlinearity, we obtain single-exponent standard scaling with $\kappa = 0$, $\alpha_\mathrm{loc} = \alpha = 2\theta +1 -d/2$, as given by Eq.~(\ref{eq:EW_exponents}).

\section{Numerical results in $d=1$}\label{sec:numeric}
For the numerical integration of the $1+1$ dimensional ALM model with time correlated noise we use an Euler scheme and discretize the equation of motion (\ref{eq:our_model}). For the site $i \in [1,N]$ we have
\begin{eqnarray*}
h_i(t+1) &=& h_i(t) + \Delta t \left[\nu \mathcal{L}_i(t) + c_{2n} \mathcal{N}^{(n)}_i(t) \right] + \\ 
&+& \Delta t \, \eta_i(t),
\label{eq:euler}
\end{eqnarray*}
where the discretized Laplacian term is
\begin{equation*}
\mathcal{L}_i(t) = a^{-2} \left[h_{i+1}(t) +  h_{i-1}(t) - 2 h_{i}(t)\right],
\end{equation*}
and the $n$-nonlinearity is given by
\begin{eqnarray*}
\mathcal{N}^{(n)}_i(t) &=& a^{-2n} \{\left[h_{i+1}(t)-h_{i}(t)\right]^{2n-1} - \\
&-& \left[h_{i}(t)-h_{i-1}(t)\right]^{2n-1}\}.
\end{eqnarray*}
The lattice spacing $a=1$, time step $\Delta = 0.01$, and periodic boundary conditions $h_{L+1}(t) = h_1(t)$ and $h_{0}(t) = h_L(t)$ are used in all our simulations. We studied systems of size $L=1024$, $2048$, and $4096$, where statistical averages over $100$ independent runs were taken for the largest system and many more for smaller ones. For sake of presentation we report here our results for $L=4096$ and the interface evolution was followed up to the stationary regime at times of the order $t = 10^4$.

The noise is Gaussian distributed and has zero mean $\langle\eta_i\rangle = 0$ and correlator given by
\begin{equation*}
\langle\eta_i(t) \eta_{i'}(t')\rangle = 2 D \, \delta_{i,i'} \, |t-t'|^{2\theta-1}.
\end{equation*}
The temporally correlated noise was generated by using the Mandelbrot fractional noise algorithm~\cite{Mandelbrot1971} that produces high quality Gaussian distributed random numbers with the desired correlation index $\theta$. This technique has proven to be very efficient   
and precise~\cite{Ales2019,Ales2020,Lam1992} for systems similar to ours. 
\begin{figure}[h]
	\centerline{\includegraphics[width=0.40\textwidth,clip]{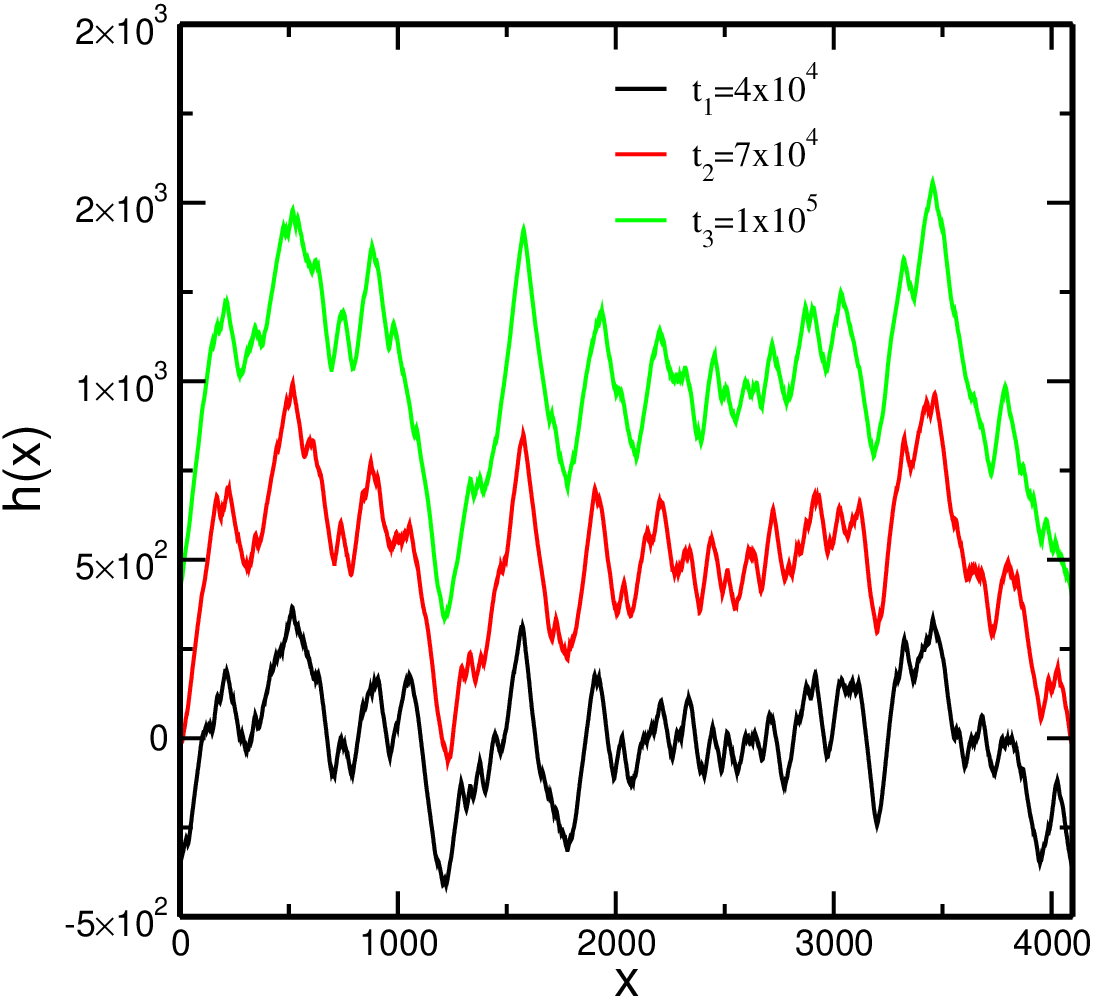}}
	\caption{Snapshots at three different times for the generalized ALM with $n=2$ in a system of size $L=4096$ and noise correlation index $\theta = 0.45$. The interface develops a `faceted' structure characterized by long faces of constant slope. Profiles are vertically shifted for easy viewing.}
	\label{fig:profiles_n2}
\end{figure}

We carried out simulations for several values of $n$ and present here our detailed results for $n = 2, 4$ and $8$. In order to speed up convergence to the true asymptotic regime we take $\nu = 0$, so that we can attain the long time limit for the moderate system sizes we use. This is very convenient since the generation of a long array of temporally correlated noise for each site requires very long computing times, which limits the system sizes that are affordable with limited computer resources. The nonlinear coupling constant is set to $c_{2n}=1$ in all cases.

\begin{figure}[h]
	\centerline{\includegraphics[width=0.40\textwidth,clip]{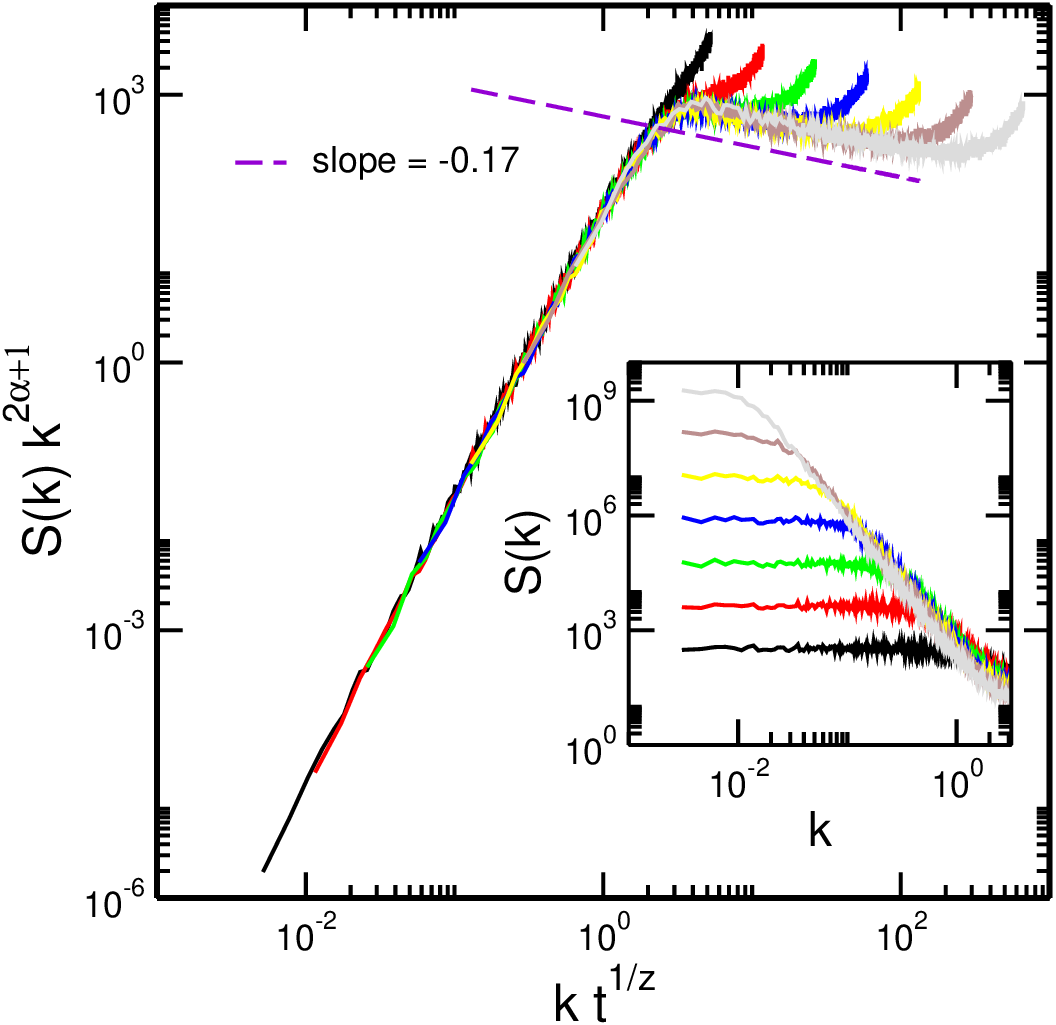}}
	\caption{Numerical calculations of the structure factor for the generalized ALM with for $n=2$, $L=4096$, and noise correlation index $\theta = 0.45$. Inset shows data collapse with the typical anomalous behavior of the scaling function at large $kt^{1/z}$ values, which decays as a power-law instead of becoming constant.}
	\label{fig:Skt_n2}
\end{figure}

In Fig.~\ref{fig:profiles_n2} we plot an example of the resulting interface profiles for $n=2$ computed at three different times for a noise correlation exponent $\theta = 0.45$, very close to the columnar disorder limit, for a system of size $L=4096$. Notice that the interface attains an apparent faceted profile with a characteristic facet base size that increases in time. This is in fact corroborated by the the structure factor $S(k,t)$ shown in the inset Fig.~\ref{fig:Skt_n2}, where a slight, but readily observable, downwards displacement for different times of the spectral densities in the large momenta region prototypical of anomalous scaling~\cite{Ramasco.etal_2000}. Indeed, spectra are nicely described by the generic ansatz, Eqs.~(\ref{eq:S}) and (\ref{eq:generic}), as shown in the main panel by a data collapse with the critical exponents $\alpha = 1.13$ and $z= 1.72$. The non zero slope shown by the collapsed spectra for $k t^{1/z} \gg 1$ immediately indicates the existence of anomalous scaling of the faceted type with $\alpha_s = 1.30 $, in agreement with the observed slope $2(\alpha - \alpha_s) \approx -0.17$. 
\begin{figure}[h]
	\centerline{\includegraphics[width=0.40\textwidth,clip]{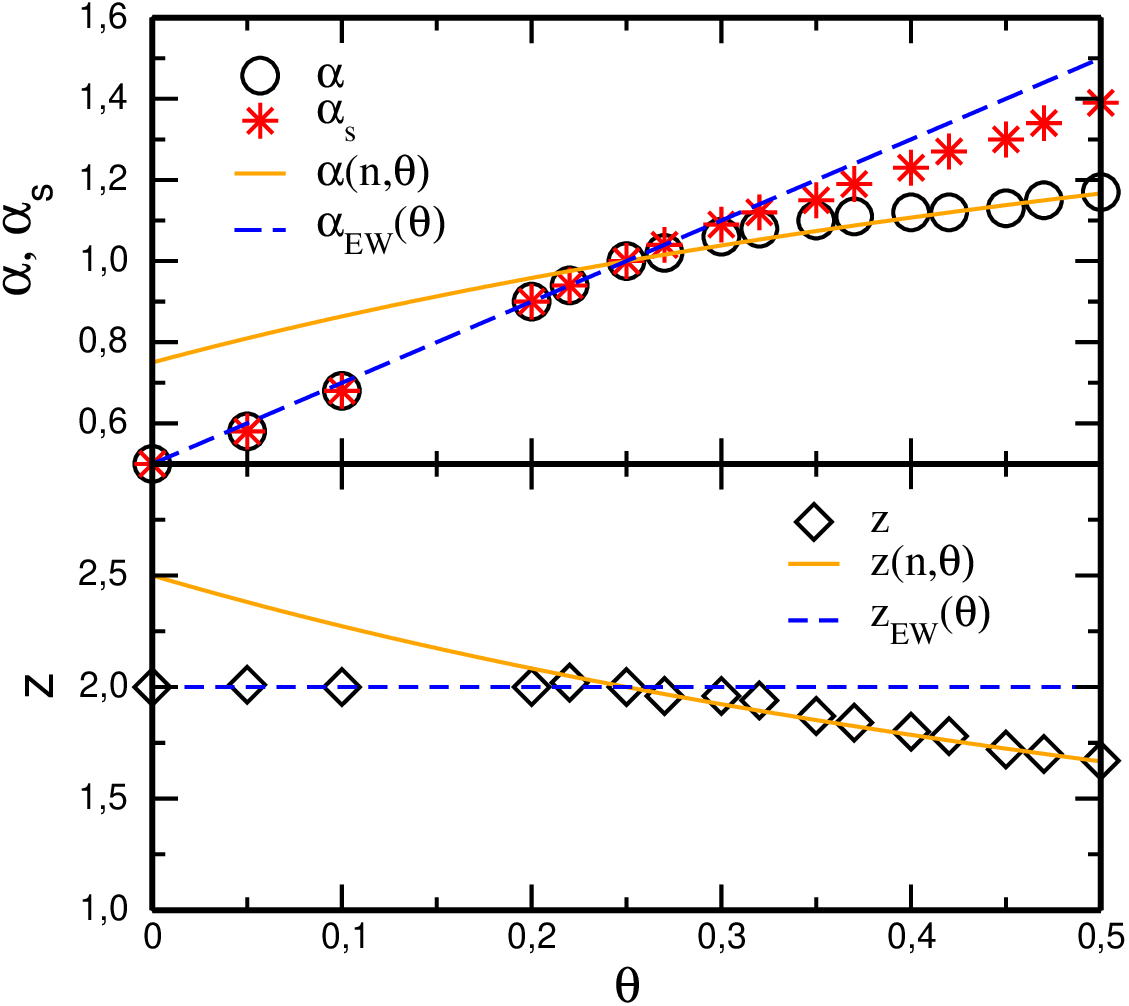}}
	\caption{Numerical estimates for the critical exponents $\alpha$, $\alpha_\mathrm{s}$, and $z$ for the generalized ALM with $n=2$ for varying noise correlation index $\theta$ are compared with the prediction of the scaling theory. Results for system size $L=4096$.}
	\label{fig:exponents_n2}
\end{figure}

We integrated the discretized equation of motion for a range of values $\theta \in [0,1/2]$, collapsed the resulting spectra, and obtained all three critical exponents $\alpha_s$, $\alpha$ and $z$. The numerical results are summarized in Fig.~\ref{fig:exponents_n2}, together with the predictions from the scaling theory for $n=2$ and $d=1$. One can observe there is a threshold $\theta_\mathrm{c} \approx 0.25$ separating two regimes. For $\theta > 0.25$, $\alpha$ and $z$ correspond to the theoretical values obtained for the nonlinear regime, Eqs.~(\ref{eq:alpha}) and (\ref{eq:z}), while for $\theta < 0.25$ the exponents are given by the Laplacian term, Eq.~(\ref{eq:z}). This numerical results are in excellent agreement with the scaling theory.
\begin{figure}[h]
	\centerline{\includegraphics[width=0.40\textwidth,clip]{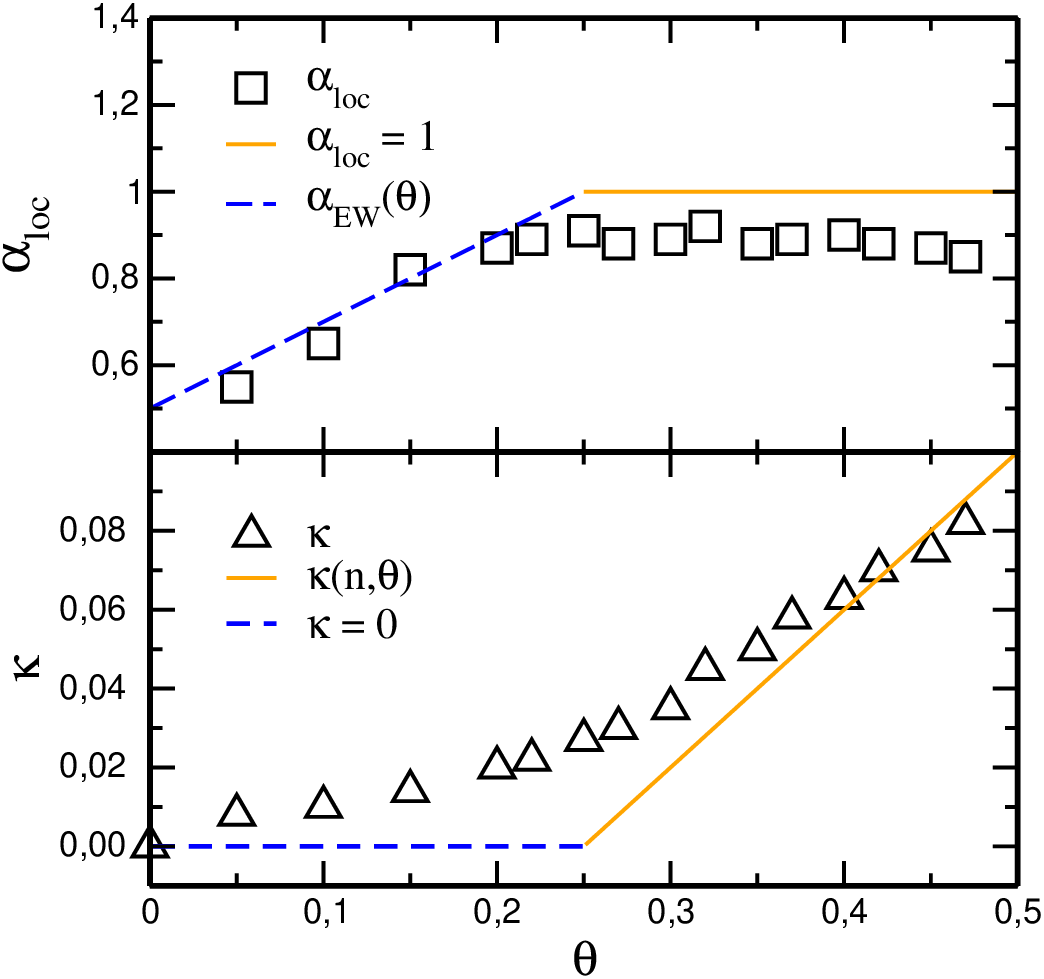}}
	\caption{Numerically obtained values for the local roughness exponent $\alpha_\mathrm{loc}$ and the anomalous time exponent $\kappa$ for $n=2$ and comparison with the the scaling theory.}
	\label{fig:local_exponents_n2}
\end{figure}

We have also measured the local roughness exponent $\alpha_\mathrm{loc}$ and anomalous time exponent $\kappa$ that describe the scaling of the height-height correlation function, $C(\ell,t) \sim \ell^{\alpha_\mathrm{loc}} \, t^\kappa$, at intermediate times $\ell \ll t^{1/z} \ll L$. In Fig.~\ref{fig:local_exponents_n2} we compare our numerical results with the scaling theory predictions in Eq.~(\ref{eq:kappa}) and (\ref{eq:alpha_loc}). The deviations of the numerically observed $\kappa$ from the prediction might be due to the difficulty in measuring precisely such a small exponent in log-log plots. In any case, the agreement of the observed local roughness exponent with theory is excellent and reassures us in the validity of our theoretical predictions. Note again the change occurring at around $\theta \approx 0.25$, as predicted by the theory, from the linear Laplacian dominated regime to the nonlinear phase.

\begin{figure}[h]
	\centerline{\includegraphics[width=0.40\textwidth,clip]{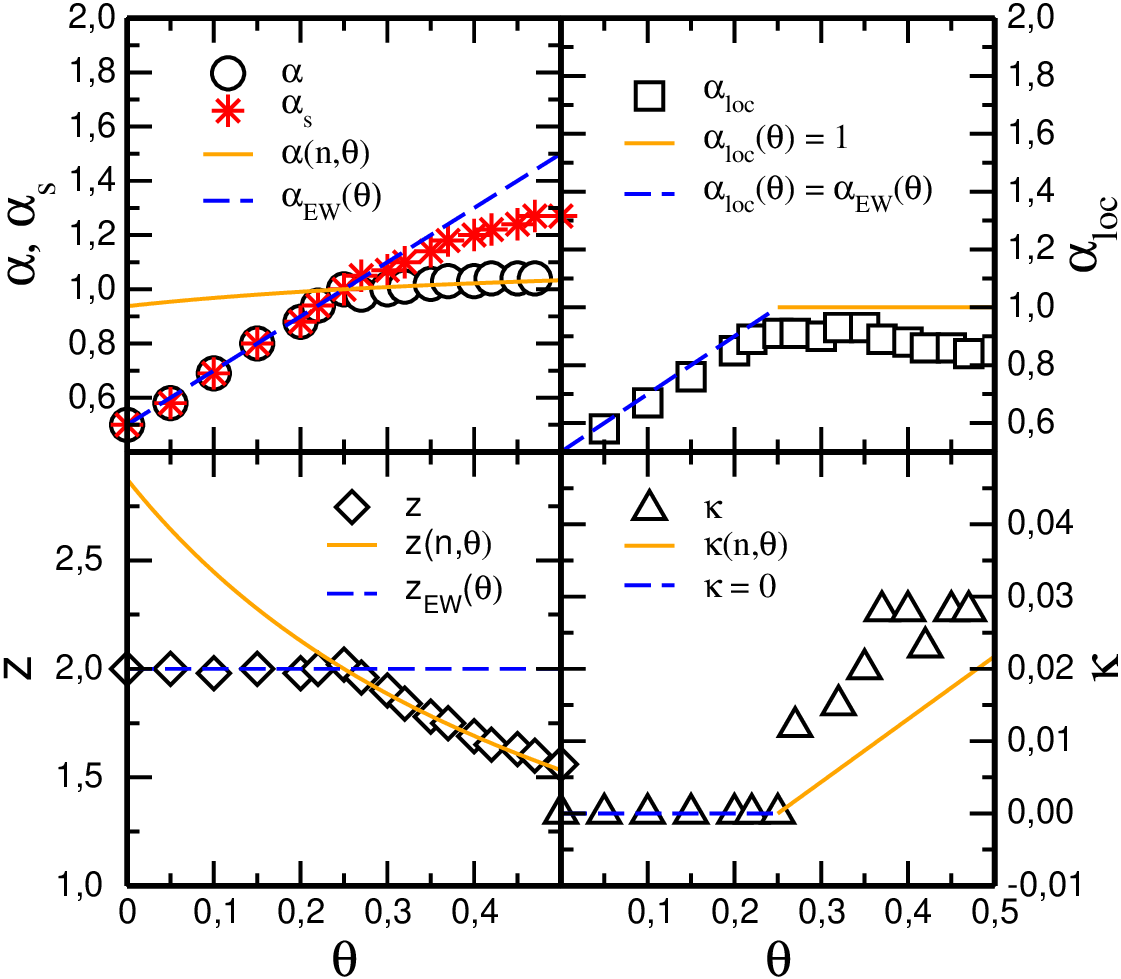}}
	\caption{Summary of all the critical exponents for the ALM with $n=8$ 
		for varying correlated noise in a system of size $L=4096$. Scaling theory predictions are also plotted for comparison.}
	\label{fig:exponents_n8}
\end{figure}

Our numerical results for nonlinearities $n=4$ and $n=8$ exhibit a similar agreement with the theoretical predictions and we just show here our results for the latter case. In Fig.~\ref{fig:exponents_n8} we summarize our numerical results for all the critical exponents and compare with theoretical predictions. For all $n$ studied there is a spectral roughness exponent $\alpha_s$ that differs from the global one $\alpha$ for $\theta > \theta_\mathrm{c} \approx 0.25$ in excellent agreement with the scaling theory, indicating that indeed the interface becomes faceted above the correlation index threshold.

\section{Discussion}\label{sec:discussion}
The Larkin model first appeared as a rough approximation to the more involved QEW by dropping the dependence of the disorder on the interface state. By doing so one arrives at a model with static columnar disorder that can be completely solved analytically. Despite its simplicity the Larkin model is a good starting point to study the effects of disorder on kinetic roughening while, at the same time, it is related to other models of interface roughening. In fact, the model can also be seen as the limit of the EW model with temporally correlated noise as $\theta \to 1/2$, which was recently studied in detail~\cite{Ales2020}. A long standing question is whether the model is physically meaningful in $d=1$ since it leads to a roughness exponent $\alpha = 3/2$ so that the elastic string is characterized by local deformations that diverge as $W(L)/L \sim L^{1/2}$ in the thermodynamic limit. While there have been many experiments in different contexts in which values of $\alpha > 1$ have been actually measured, the important point here is whether the elastic approximation that gives rise to the Larkin model, after neglecting higher-order terms in the expansion of the elastic energy, makes sense or should be corrected. Actually, similar concerns can be risen for the cases of related models like the just mentioned EW model with temporally correlated noise or the QEW model, both coming from the harmonic approximation to the elastic Hamiltonian and exhibiting $\alpha > 1$. Interestingly, it was shown~\cite{Rosso.Krauth_2001} that adding higher-order anharmonic terms to the QEW equation produces interfaces with roughness exponent $\alpha = 0.63$, suggesting that these corrections, when taken into account, lead to quenched KPZ behavior. Working by analogy, this result pushed Purrello {\it et al.}~\cite{Purrello2019} to investigate the effect of anharmonic corrections on the simpler Larkin model, the so-called ALM. These authors concluded that, far from curing the problem, the ALM in $d=1$ leads to faceted interfaces with the corresponding anomalous scaling characterized by a spectral roughness exponent $\alpha_s > \alpha > 1$ for any degree of anharmonicity $n$, and asked the question whether this faceted dynamics would take place in $d>1$. Here we have proven analytically that anomalous faceted scaling of the ALM is restricted to $d=1$. We have introduced a generalization of the ALM to include noise with long temporal correlations characterized by an index $\theta$ and calculated the critical exponents as a function of the anharmonicity degree $n$ and $\theta$. This has allowed us to show that anharmonic terms of order $n$ are irrelevant in the long wavelength limit for $d>1$, no matter the values of $n$ or $\theta$, due to the fact that kinetic roughening is dominated by the Laplacian term. Only for $d=1$ and correlation index larger than $\theta_\mathrm{c} = 1/4$ the anharmonic terms become relevant, the interface forms facets with $\alpha_s \neq \alpha$, and anomalous scaling occurs. This includes, as a particular instance, the case treated by Purrello {\it et al.}~\cite{Purrello2019}, that is $\theta = 1/2$ in $d=1$.

Finally, it is interesting to remark that the effects of time correlated noise in the ALM are similar to those recently observed in other interface models, like the EW and the KPZ systems. Specifically, all models share in common the existence of a threshold value $\theta_\mathrm{c} = 1/4$ for anomalous behavior, which seems to be robust across different universality classes in $d=1$. For EW, the interface shows anomalous scaling of the super-rough (non faceted) type~\cite{Ales2020} with $\alpha_s(\theta) = \alpha(\theta) > 1$ and $\alpha_\mathrm{loc} = 1$, while for KPZ the scaling is anomalous faceted type~\cite{Ales2019}, with $\alpha_s(\theta) \neq \alpha(\theta)$ and $\alpha_\mathrm{loc} = 1$, in both cases for $\theta > 1/4$. The addition of anharmonic corrections to the EW equation studied here produces anomalous faceted scaling for a noise correlation index $\theta > \theta_\mathrm{c} = d/4$ suggesting that facet formation is a nonlinear effect. Indeed, no linear model is known to be able to produce facets. 

\acknowledgments
This work has been partially supported by the Program for Scientific Cooperation
I-COOP+ from Consejo Superior de Investigaciones Científicas (Spain) through project No. COOPA20187. AA thanks financial support from Programa de
Pasant\'ias de la Universidad de Cantabria in 2017 and 2018 (Projects No. 70-ZCE3-
226.90 and 62-VCES-648), and CONICET (Argentina) for a post-doctoral fellowship.
JML is partially supported by project No. FIS2016-74957-P from Agencia Estatal de
Investigación (AEI) and FEDER (EU).

\bibliographystyle{prsty}

\begin{thebibliography}{10}
	
	\bibitem{Barabasi.Stanley1995_book}
	A.-L. Barab{\'a}si and H.~E. Stanley, {\em Fractal Concepts in Surface Growth}
	(Cambridge University Press, Cambridge, 1995), p.\ 386.
	
	\bibitem{Krug1997}
	J. Krug, Adv. Phys. {\bf 46},  139  (1997).
	
	\bibitem{Fisher1998}
	D. Fisher, Phys. Rep. {\bf 301},  113  (1998).
	
	\bibitem{Kardar_1998}
	M. Kardar, Phys. Rep. {\bf 301},  85   (1998).
	
	\bibitem{Nattermann1992}
	T. Nattermann, S. Stepanow, L.-H. Tang, and H. Leschhorn, J. Phys. II France
	{\bf 2},  1483  (1992).
	
	\bibitem{Leschhorn1997}
	H. Leschhorn, T. Nattermann, S. Stepanow, and L.-H. Tang, Ann. Phys. {\bf 509},
	1  (1997).
	
	\bibitem{Jensen_1995}
	H.~J. Jensen, J. Phys. A: Math. Gen. {\bf 28},  1861  (1995).
	
	\bibitem{Rosso.Krauth_2001}
	A. Rosso and W. Krauth, Phys. Rev. B {\bf 65},  012202  (2001).
	
	\bibitem{Duemmer.Krauth_2005}
	O. Duemmer and W. Krauth, Phys. Rev. E {\bf 71},  061601  (2005).
	
	\bibitem{Rosso.etal_2003}
	A. Rosso, A.~K. Hartmann, and W. Krauth, Phys. Rev. E {\bf 67},  021602
	(2003).
	
	\bibitem{Rosso.Krauth_2002}
	A. Rosso and W. Krauth, Phys. Rev. E {\bf 65},  025101  (2002).
	
	\bibitem{Chauve.etal_2001}
	P. Chauve, P. Le~Doussal, and K. J\"org~Wiese, Phys. Rev. Lett. {\bf 86},  1785
	(2001).
	
	\bibitem{Kolton.etal_2005a}
	A.~B. Kolton, A. Rosso, and T. Giamarchi, Phys. Rev. Lett. {\bf 95},  180604
	(2005).
	
	\bibitem{Bruinsma.Aeppli_1984}
	R. Bruinsma and G. Aeppli, Phys. Rev. Lett. {\bf 52},  1547  (1984).
	
	\bibitem{Halpin-Healy_1989}
	T. Halpin-Healy, Phys. Rev. Lett. {\bf 62},  442  (1989).
	
	\bibitem{Huse.Henley_1985}
	D.~A. Huse and C.~L. Henley, Phys. Rev. Lett. {\bf 54},  2708  (1985).
	
	\bibitem{Blatter1994}
	G. Blatter {\it et~al.}, Rev. Mod. Phys. {\bf 66},  1125  (1994).
	
	\bibitem{Ertas_1996}
	D. Erta\ifmmode~\mbox{\c{s}}\else \c{s}\fi{} and M. Kardar, Phys. Rev. B {\bf
		53},  3520  (1996).
	
	\bibitem{Nattermann2000}
	T. Nattermann and S. Scheidl, Adv. Phys. {\bf 49},  607  (2000).
	
	\bibitem{Ramanathan.etal_1997}
	S. Ramanathan, D. Erta\c{s}, and D.~S. Fisher, Phys. Rev. Lett. {\bf 79},  873
	(1997).
	
	\bibitem{Bouchaud1993}
	J.~P. Bouchaud, E. Bouchaud, G. Lapasset, and J. Plan\`es, Phys. Rev. Lett.
	{\bf 71},  2240  (1993).
	
	\bibitem{Bouchaud_1997}
	E. Bouchaud, J. Phys. Condens. Matter {\bf 9},  4319  (1997).
	
	\bibitem{Ponson2017}
	L. Ponson and N. Pindra, Phys. Rev. E {\bf 95},  053004  (2017).
	
	\bibitem{Dube1999}
	M. Dub\'e {\it et~al.}, Phys. Rev. Lett. {\bf 83},  1628  (1999).
	
	\bibitem{Hernandez-MachadoA.2001}
	{Hern\'andez-Machado, A.} {\it et~al.}, Europhys. Lett. {\bf 55},  194  (2001).
	
	\bibitem{Alava.etal_2004}
	M. Alava, M. Dub{\'e}, and M. Rost, Adv. Phys. {\bf 53},  83  (2004).
	
	\bibitem{Rost.etal_2007}
	M. Rost, L. Laurson, M. Dub{\'e}, and M. Alava, Phys. Rev. Lett. {\bf 98},
	054502  (2007).
	
	\bibitem{Pradas2007}
	M. Pradas, J.~M. L\'opez, and A. Hern\'andez-Machado, Phys. Rev. E {\bf 76},
	010102  (2007).
	
	\bibitem{Pradas2009}
	M. Pradas, J.~M. L\'opez, and A. Hern\'andez-Machado, Phys. Rev. E {\bf 80},
	050101  (2009).
	
	\bibitem{Nattermann.etal_1990}
	T. Nattermann, Y. Shapir, and I. Vilfan, Phys. Rev. B {\bf 42},  8577  (1990).
	
	\bibitem{Chauve2000}
	P. Chauve, T. Giamarchi, and P. Le~Doussal, Phys. Rev. B {\bf 62},  6241
	(2000).
	
	\bibitem{Kolton2009}
	A.~B. Kolton, A. Rosso, T. Giamarchi, and W. Krauth, Phys. Rev. B {\bf 79},
	184207  (2009).
	
	\bibitem{de_la_Lama_2009}
	M.~S. de~la Lama, J.~M. L{\'{o}}pez, J.~J. Ramasco, and M.~A. Rodr{\'{\i}}guez,
	J. Stat. Mech.: Theory Exp {\bf 2009},  P07009  (2009).
	
	\bibitem{Purrello2017}
	V.~H. Purrello, J.~L. Iguain, A.~B. Kolton, and E.~A. Jagla, Phys. Rev. E {\bf
		96},  022112  (2017).
	
	\bibitem{Makse_1995}
	H.~A. Makse and L.~A.~N. Amaral, Europhysics Letters ({EPL}) {\bf 31},  379
	(1995).
	
	\bibitem{Amaral1995}
	L.~A.~N. Amaral, A.-L. Barab\'asi, H.~A. Makse, and H.~E. Stanley, Phys. Rev. E
	{\bf 52},  4087  (1995).
	
	\bibitem{Larkin1979}
	A. Larkin and Y. Ovchinnikov, J. Low. Temp. Phys. {\bf 34},  409–428  (1979).
	
	\bibitem{Zhang1986}
	Y.~C. Zhang, Phys. Rev. Lett. {\bf 56},  2113  (1986).
	
	\bibitem{Engel1987}
	A. Engel and W. Ebeling, Phys. Rev. Lett. {\bf 59},  1979  (1987).
	
	\bibitem{Tao1988}
	R. Tao, Phys. Rev. Lett. {\bf 61},  2405  (1988).
	
	\bibitem{Guyer1990}
	R.~A. Guyer and J. Machta, Phys. Rev. Lett. {\bf 64},  494  (1990).
	
	\bibitem{Szendro2007a}
	I.~G. Szendro, J.~M. L\'opez, and M.~A. Rodr\'{\i}guez, Phys. Rev. E {\bf 76},
	011603  (2007).
	
	\bibitem{Ramasco.etal_2000}
	J.~J. Ramasco, J.~M. L\'{o}pez, and M.~A. Rodr\'{i}guez, Phys. Rev. Lett. {\bf
		84},  2199  (2000).
	
	\bibitem{footnote} Note that the Taylor expansion for this particular choice of the elastic energy leads to terms with alternating signs. This is irrelevant for the scaling theory but not for simulations. Given that we will be taking the diffusion coefficient $\nu=0$ to speed up the transient to the asymptotic regime in our numerical integration, all coefficients $c_{2n}$ must be positive to assure numerical stability. This assumption is in the line of previous studies of these type of anharmonic corrections in the similar systems~\cite{Rosso.Krauth_2001,Purrello2019}.
	
	\bibitem{Leschhorn1993}
	H. Leschhorn, Physica A {\bf 195},  324  (1993).
	
	\bibitem{Rosso2001}
	A. Rosso and W. Krauth, Phys. Rev. Lett. {\bf 87},  187002  (2001).
	
	\bibitem{Tang.Leschhorn_1992}
	L.-H. Tang and H. Leschhorn, Phys. Rev. A {\bf 45},  R8309  (1992).
	
	\bibitem{Buldyrev1992}
	S.~V. Buldyrev {\it et~al.}, Phys. Rev. A {\bf 45},  R8313  (1992).
	
	\bibitem{Purrello2019}
	V.~H. Purrello, J.~L. Iguain, and A.~B. Kolton, Phys. Rev. E {\bf 99},  032105
	(2019).
	
	\bibitem{Lopez_1999}
	J.~M. L\'{o}pez, Phys. Rev. Lett. {\bf 83},  4594  (1999).
	
	\bibitem{Ales2020}
	A. Al{\'{e}}s and J.~M. L{\'{o}}pez, J. Stat. Mech.: Theory Exp {\bf 2020},
	033210  (2020).
	
	\bibitem{Mandelbrot1971}
	B.~B. Mandelbrot, Water Resour. Res. {\bf 7},  543  (1971).
	
	\bibitem{Ales2019}
	A. Al\'es and J.~M. L\'opez, Phys. Rev. E {\bf 99},  062139  (2019).
	
	\bibitem{Lam1992}
	C.-H. Lam, L.~M. Sander, and D.~E. Wolf, Phys. Rev. A {\bf 46},  R6128  (1992).
	
\end{thebibliography}

\end{document}